\documentclass[reprint,amsmath,amssymb,aps,prl,superscriptaddress,nofootinbib]{revtex4-2}
\usepackage[T1]{fontenc}
\usepackage{graphicx}
\usepackage{dcolumn}
\usepackage{bm}
\usepackage{mathrsfs}
\usepackage{amsmath}
\usepackage{physics}
\usepackage{hyperref}
\usepackage{latexsym}
\usepackage{array}
\usepackage{color}
\usepackage{multirow}
\usepackage{booktabs}

\hypersetup{hypertex=true, colorlinks=true, linkcolor=blue, urlcolor=blue, citecolor=magenta}

\begin{document}

\title{Electrostatic-Elastic Softening and Ultraviolet Instability Driven by Non-DLVO Interactions in Charged Colloidal Crystals}
\author{Hao Wu}
\email{wuhao@ucas.ac.cn}
\affiliation{Zhejiang Key Laboratory of Soft Matter Biomedical Materials, Wenzhou Institute, University of Chinese Academy of Sciences, Wenzhou, Zhejiang 325000, China;}
\author{Zhong-Can Ou-Yang}
\affiliation{Institute of Theoretical Physics, Chinese Academy of Sciences, Beijing 100190, China.}

\date{\today}

\begin{abstract}
Colloidal crystals permeated by mobile ions exhibit a coupling between electrostatic and elastic degrees of freedom that renormalizes the effective screening length and induces wave-vector-dependent elastic softening. Building on our recently proposed continuum model, we perform a rigorous Gaussian fluctuation analysis to elucidate the stability limits of the homogeneous phase. By integrating out the electrostatic fluctuations, we derive the effective elastic modulus $\Gamma(q)$ as a function of wave vector $q$. We show that the modulus in the long-wavelength limit ($q \to 0$) remains identically equal to a bare modulus protected by perfect ionic screening. In contrast, the modulus in the short-wavelength limit ($q \to \infty$) softens as the electrostatic-elastic coupling strength $\xi$ increases, vanishing at a critical value $\xi=1$. For $\xi>1$, the fluctuation spectrum exhibits a negative eigenvalue for all wave vectors $q$ larger than a critical (effective screening) wave vector $q_c$, signaling an ultraviolet instability of the uniform phase. In a real colloidal crystal, this divergence is regulated by the discrete lattice cutoff $q_{\max}\sim\pi/a$, confining the physical instability to a finite band $q_c < q < q_{\max}$. The macroscopic limit $q\to 0$ remains unconditionally stable for all $\xi$. The transition at $\xi=1$ thus marks the onset of short-wavelength mechanical failure, while macroscopic elastic stiffness remains intact. Our analysis clarifies the proper physical interpretation of the minimal coupling model and provides a consistent picture of how non-DLVO interactions can drive local structural collapse in charged colloidal crystals.
\end{abstract}

\maketitle

\section{Introduction}

In classical atomic crystals, the~coupling between the quantum behavior of electrons and lattice vibrations leads to extraordinary macroscopic quantum phenomena such as Cooper pairing and superconductivity. In~the entirely classical realm of physics, colloidal crystals—ordered arrays of charged large colloidal particles with mobile small ions occupying the interstitial spaces—offer an experimental platform to observe “electron-phonon-like” coupling effects at room temperature~\cite{Girard2019,Lechner2009a}. In~recent years, with~the development of binary colloidal crystals and DNA-functionalized nanoparticle systems, researchers have discovered that interstitial ions can undergo a continuous transition from localized to delocalized states, a~phenomenon analogized as an “ionic-to-metallic” transition~\cite{Lin2022}, sparking widespread discussion on the nature of effective interactions in classical~systems.

A fundamental distinction between colloidal and atomic crystals lies in their classical statistical mechanics: quantum fluctuations are replaced by thermal fluctuations, and~quantum exchange is replaced by classical electrostatics. This raises a profound theoretical question: When classical electrostatics couples with lattice elasticity, how do fluctuations affect the mechanical stability of the system? This question not only concerns the foundations of colloid science but also attracts broad interest due to its connections to defect interactions~\cite{Lechner2008,Lechner2009b}, critical Casimir forces~\cite{Maciolek2018,Gambassi2024}, and~soft matter self-assembly~\cite{Ciach2013,Ciach2009}.

Beyond the conventional Derjaguin-Landau-Verwey-Overbeek (DLVO) paradigm~\cite{Derjaguin1941,Verwey1948}, there exists an “embarras de richesses” of non-DLVO interactions in colloidal systems~\cite{Podgornik2021}, among~which lattice-elasticity-mediated interactions constitute an important branch. Recent experiments have provided direct evidence for attractive non-DLVO forces between like-charged colloidal particles~\cite{Smith2017}, and~the physics of strongly coupled Coulomb fluids has been systematically reviewed~\cite{Boroudjerdi2005}. These developments underscore the need for a deeper understanding of how electrostatic and elastic degrees of freedom conspire to produce novel collective~phenomena.

Wu~et~al. recently proposed an elegant minimal coupling continuum model that directly addresses this interplay~\cite{Wu2025,Wu2024EPL}. They coupled Poisson–Boltzmann electrostatics with Hookean elasticity via a local dilation term. Mean-field analysis showed that the electrostatic--elastic coupling renormalizes the screening length, yielding an effective screening wave-vector squared $\kappa_{\text{eff}}^2 = \kappa_0^2/(1-\xi)$, where the dimensionless parameter $\xi \equiv 2\beta n_0 v_0^2 K$ measures the coupling strength. A~striking result emerges: as $\xi \to 1^-$, $\kappa_{\text{eff}} \to \infty$, i.e.,~the screening length tends to zero—the stronger the coupling, the~more localized the screening. This counterintuitive “critical collapse” strongly suggests that at $\xi=1$ the uniform bulk phase must face some kind of instability. This discovery has also inspired further studies on anisotropic colloidal crystals~\cite{Wu2026anisotropic}.

{To place our continuum model in a proper physical context, it is instructive to specify the relevant length scales of the typical colloidal crystals considered here. In~such systems, the~macroscopic charged colloidal particles typically have radii $a$ ranging from $50\,\mathrm{nm}$ to $1\,\mathrm{\mu m}$, while the average lattice spacing $d$ between them spans from hundreds of nanometers to several micrometers~\cite{Yethiraj2007,Hansen2000,Crocker1994}. Because~the wavelength of the elastic deformation we consider is much larger than $d$, the~continuum description for the elastic displacement field is well justified. Furthermore, the~surrounding mobile co-ions and counterions have sub-nanometer hydrated radii ($r_{\mathrm{ion}} \sim 0.3\,\mathrm{nm}$). Given the profound scale separation ($r_{\mathrm{ion}} \ll a < d$), it is a standard and robust approximation to treat these mobile ions as point-like charges~\cite{Levin2002}. This point-ion assumption underpins the classical Poisson--Boltzmann functional used in our field-theoretic formulation, focusing our attention on the long-wavelength electrostatic--elastic coupling rather than ionic excluded-volume effects.}

The purpose of this paper is to perform a rigorous Gaussian fluctuation analysis of this model and to clarify the nature of the instability at $\xi=1$. We shall demonstrate~that:
\begin{enumerate}
    \item The effective long-wavelength bulk modulus $\Gamma(0)$ remains exactly equal to the bare modulus $\beta K$, protected by perfect ionic screening at macroscopic scales.
    \item The effective short-wavelength modulus $\Gamma(q\to\infty) = \beta K(1-\xi)$ softens as $\xi$ increases and vanishes at $\xi=1$, signaling a local mechanical instability at small scales.
    \item The fluctuation spectrum reveals that for $\xi>1$ the uniform phase becomes unstable against perturbations with wave vectors $q>q_c = \kappa_0/\sqrt{\xi-1}$. In~the minimal model, this instability extends to arbitrarily short wavelengths, a~feature that must be regulated by the discrete lattice cutoff in real systems.
    \item The effective screening wave vector diverges as $\xi\to1$, and~for $\xi>1$ it becomes purely imaginary, indicating a crossover from exponential screening to a structural instability driven by the dominance of electrostatic attraction over elastic restoring~forces.
\end{enumerate}

Our analysis shows that within~the minimal coupling model, the~transition at $\xi=1$ corresponds to a short-wavelength electrostatic-elastic spinodal. Although~higher-order strain-gradient terms are required to describe a true modulated phase, the~present work provides a quantitative description of the instability threshold and clarifies the proper physical interpretation of the renormalized elastic~moduli.

\section{Model System and Field-Theoretic~Formulation}
\unskip

\subsection{From Microscopic Degrees of Freedom to a Continuum~Action}

Consider a three-dimensional colloidal crystal with a background ideal lattice of large charged colloidal particles, permeated by mobile monovalent positive and negative ions. In~the continuum approximation, the~colloidal lattice is regarded as an isotropic elastic medium whose elastic distortion is described by the displacement field $\mathbf{u}(\mathbf{x})$; the mobile ions form a Coulomb fluid whose electrostatic properties are characterized by the mean electrostatic potential $\phi(\mathbf{x})$. The~total free energy of the system (divided by $k_B T$ to obtain a dimensionless action) can be written as~\cite{Wu2025}:
\begin{align}
\label{eq:total_action}
S[\phi, \mathbf{u}] = \beta \mathcal{F} =& \int \dd^3 x \Big[ -\frac{\beta \epsilon}{2} (\nabla \phi)^2 + \frac{\beta K}{2} (\nabla \cdot \mathbf{u})^2 \notag\\
&- 2 n_0 \cosh\big( \beta e_0 \phi + \beta v_0 K \nabla \cdot \mathbf{u} \big) \Big].
\end{align}
\indent{The} symbols have the following physical~meanings:
\begin{itemize}
    \item $\epsilon = \epsilon_r \epsilon_0$: dielectric permittivity of the solvent;
    \item $K = \lambda + 2\mu$: bulk modulus of the colloidal lattice ($\lambda, \mu$ are the Lam\'e coefficients), which can be computed from Poisson--Boltzmann based models~\cite{Dyshlovenko2010};
    \item $n_0$: bulk number density of ions far from interfaces;
    \item $\beta = 1/(k_B T)$: inverse temperature;
    \item $e_0$: elementary charge, whose effective value can be determined from conductivity and shear modulus measurements~\cite{Wette2003};
    \item $v_0$: local volume change induced by a single interstitial charge, characterizing the ``elastic charge'' strength.
\end{itemize}

In Equation \eqref{eq:total_action}, the~first term is the electrostatic field energy (the negative sign arises from the structure of standard PB field theory and can be handled via a Wick rotation to imaginary time), the~second term is the isotropic elastic energy, and~the third term is the ideal-gas entropy of the mobile ions, with~the hyperbolic cosine representing the sum of Boltzmann factors for positive and negative~ions.

\subsection{Composite Field and Scalar~Reduction}

Notice that the elastic displacement field $\mathbf{u}$ enters the free energy only through its divergence $\nabla \cdot \mathbf{u}$. Under~the minimal coupling assumption, only isotropic volume expansion/contraction couples to the electrostatic potential, while shear strains decouple. Hence we introduce a scalar field $\theta(\mathbf{x})$ to describe the local volumetric strain:
\begin{equation}
\theta(\mathbf{x}) \equiv \nabla \cdot \mathbf{u}(\mathbf{x}).
\end{equation}
Furthermore, we define two fundamental coupling constants:
\begin{align}
g_1 &\equiv \beta e_0, \label{eq:def_g1} \\
g_2 &\equiv \beta v_0 K, \label{eq:def_g2}
\end{align}
and construct a composite scalar field $\Phi(\mathbf{x})$:
\begin{equation}\label{eq:def_Phi}
\Phi(\mathbf{x}) \equiv g_1 \phi(\mathbf{x}) + g_2 \theta(\mathbf{x}).
\end{equation}
Substituting into the action \eqref{eq:total_action}, we obtain its compact form:
\begin{equation}\label{eq:S_compact}
S = \int \dd^3 x \left[ -\frac{\beta \epsilon}{2} (\nabla \phi)^2 + \frac{\beta K}{2} \theta^2 - 2 n_0 \cosh(\Phi) \right].
\end{equation}
This form clearly shows that all nonlinear interactions of the system are carried solely by the single composite field $\Phi$.

\section{Gaussian Fluctuations and Effective Elastic~Modulus}
\unskip

\subsection{Mean-Field Solution and Quadratic~Expansion}

In the bulk phase, the~mean-field equations $\delta S/\delta \phi = 0$ and $\delta S/\delta \theta = 0$ admit the trivial solution $\phi_{\text{MF}}=0$, $\theta_{\text{MF}}=0$, corresponding to an electroneutral, unstrained state. Expanding $\cosh(\Phi)$ to second order around this state, we obtain the Gaussian fluctuation~action:
\begin{equation}\label{eq:S_Gauss}
\begin{aligned}
S_G = \int \dd^3 x \Big[ &-\frac{\beta \epsilon}{2} (\nabla \delta \phi)^2 + \frac{\beta K}{2} (\delta \theta)^2 \\
&- n_0 (g_1 \delta \phi + g_2 \delta \theta)^2 \Big].
\end{aligned}
\end{equation}

\subsection{Quadratic Form in Momentum~Space}

Transforming into Fourier space with $f(\mathbf{x}) = \int \frac{\dd^3 q}{(2\pi)^3} e^{i\mathbf{q}\cdot\mathbf{x}} \tilde{f}(\mathbf{q})$, the~quadratic form is $\frac{1}{2}\int_{\mathbf{q}} \mathbf{v}^\dagger \mathbf{M}(\mathbf{q})\mathbf{v}$ with $\mathbf{v}=(\delta\tilde{\phi},\delta\tilde{\theta})^T$ and
\begin{equation}\label{eq:M_matrix}
\mathbf{M}(\mathbf{q}) = \begin{pmatrix}
-\beta \epsilon q^2 - 2 n_0 g_1^2 & -2 n_0 g_1 g_2 \\
-2 n_0 g_1 g_2 & \beta K - 2 n_0 g_2^2
\end{pmatrix}.
\end{equation}

\subsection{Effective Elastic Modulus $\Gamma(q)$}

To isolate the effective elastic response, we integrate out the electrostatic fluctuations $\delta\tilde{\phi}$. This yields a purely elastic effective action for $\delta\tilde{\theta}$ with a wave-vector-dependent modulus $\Gamma(q)$:
\begin{equation}\label{eq:Gamma_def}
S_{\text{eff}}[\delta\tilde{\theta}] = \frac{1}{2} \int \frac{\dd^3 q}{(2\pi)^3} \Gamma(q) |\delta\tilde{\theta}(\mathbf{q})|^2,
\end{equation}
{By `integrating out' the electrostatic fluctuations, we mean performing the Gaussian functional integral over $\delta \phi$ in the partition function, which yields an effective action for $\delta \theta$ with a wave-vector-dependent elastic modulus $\Gamma(q)$. At~the Gaussian level, this procedure is equivalent to minimizing the quadratic action with respect to $\delta \phi$ at fixed $\delta \theta$, followed by taking into account the fluctuation determinant,}
where
\begin{equation}\label{eq:Gamma_q}
\Gamma(q) = M_{22} - \frac{M_{12}^2}{M_{11}} = \beta K - 2n_0g_2^2 + \frac{4n_0^2 g_1^2 g_2^2}{\beta\epsilon q^2 + 2n_0g_1^2}.
\end{equation}
Introducing the parameters $a = 2n_0g_1^2$, $b = \beta K - 2n_0g_2^2 = \beta K(1-\xi)$, and~$c = 2n_0g_1g_2$, we can rewrite
\begin{equation}
\Gamma(q) = b + \frac{c^2}{\beta\epsilon q^2 + a}.
\end{equation}
The physical meaning of $\Gamma(q)$ is~transparent:
\begin{itemize}
    \item Long-wavelength limit ($q\to 0$): $\Gamma(0) = b + \frac{c^2}{a} = \beta K(1-\xi) + \beta K\xi = \beta K$. The~macroscopic bulk modulus is completely unaffected by the electrostatic--elastic coupling because mobile ions have ample space to redistribute and perfectly screen the elastic deformation.
    \item Short-wavelength limit ($q\to\infty$): $\Gamma(\infty) = b = \beta K(1-\xi)$. In~the short-wavelength limit, the~induced electrostatic potential varies on a scale much finer than the Debye length $\kappa_0^{-1}$, rendering the ionic cloud spatially incapable of neutralizing the deformation-induced effective charges. Consequently, the~local elastic stiffness is reduced by the factor $(1-\xi)$.
\end{itemize}

Thus, the~electrostatic--elastic coupling induces a wave-vector-dependent softening that exclusively affects short-wavelength deformations, while the macroscopic elastic modulus remains~rigid.

\section{Eigenvalue Analysis and Instability~Threshold}

Although $\Gamma(q)$ is the proper physical modulus for the elastic sector, the~full stability of the system is determined by the eigenvalues of the original matrix $\mathbf{M}(\mathbf{q})$. Using the parameters $a,b,c$ defined above, the~matrix is
\begin{equation}
\mathbf{M}(\mathbf{q}) = \begin{pmatrix}
-\beta \epsilon q^2 - a & -c \\
-c & b
\end{pmatrix}.
\end{equation}
Its determinant is
\begin{equation}\label{eq:det_simplified}
\det \mathbf{M}(\mathbf{q}) = -\beta \epsilon b q^2 - a b - c^2.
\end{equation}
The eigenvalues are
\begin{widetext}
\begin{equation}
\lambda_{\pm}(\mathbf{q}) = \frac{-\beta \epsilon q^2 - a + b \pm \sqrt{(-\beta \epsilon q^2 - a + b)^2 - 4(-\beta \epsilon b q^2 - a b - c^2)}}{2}.
\end{equation}
\end{widetext}

\subsection{Stability for $\xi<1$}

For $\xi<1$, we have $b>0$. At~$q=0$, $\det\mathbf{M}(0) = -ab - c^2 < 0$, so one eigenvalue is always negative—a well-known feature of the PB action that is handled by Wick rotation. The~physical stability is governed by the eigenvalue that can approach zero. For~$\xi<1$, this eigenvalue remains positive for all $q$, and~the uniform phase is locally~stable.

\subsection{Onset of Instability at $\xi=1$}

As $\xi \to 1^-$, the~short-wavelength modulus $b = \beta K(1-\xi)$ tends to zero. At~$\xi=1$, \mbox{$b=0$}, and~the determinant becomes $\det\mathbf{M}(q) = -c^2 < 0$ independent of $q$. The~zero eigenvalue condition $\det\mathbf{M}=0$ cannot be satisfied for any real $q$; instead, the~system reaches a marginal stability limit where the short-wavelength stiffness vanishes. Formally, the~instability first appears at $q\to\infty$, indicating an ultraviolet divergence in the continuum~model.

\subsection{Ultraviolet Instability for $\xi>$ 1 and Lattice~Cutoff}

For $\xi>1$, $b$ becomes negative. Setting $\det\mathbf{M}=0$ gives the condition for a zero~eigenvalue:
\begin{equation}
q^2 = -\frac{a b + c^2}{\beta \epsilon b} = \frac{\kappa_0^2}{\xi - 1} \equiv q_c^2,
\end{equation}
where $\kappa_0^2 = 2\beta n_0 e_0^2/\epsilon$. Thus, for~all wave vectors $q > q_c$, the~determinant is negative and the eigenvalue $\lambda_-(\mathbf{q})$ is negative, meaning the uniform phase is unstable against perturbations with wavelengths shorter than $2\pi/q_c$. In~the minimal continuum model, this instability extends to arbitrarily large $q$, leading to an ultraviolet divergence. In~a real colloidal crystal, the~discrete lattice provides a natural ultraviolet cutoff $q_{\max} \sim \pi/a$, where $a$ is the lattice spacing. Therefore, the~actual unstable modes are restricted to the finite band $q_c < q < q_{\max}$. Note that the instability only becomes physically manifest when $\xi$ is large enough such that $q_c < q_{\max}$. The~transition at $\xi=1$ thus marks the onset of a short-wavelength mechanical instability, which in practice may result in local restructuring or microphase separation, while the macroscopic elastic modulus remains intact.

\begin{figure}[tbp]
    \centering
    \includegraphics[width=0.48\textwidth]{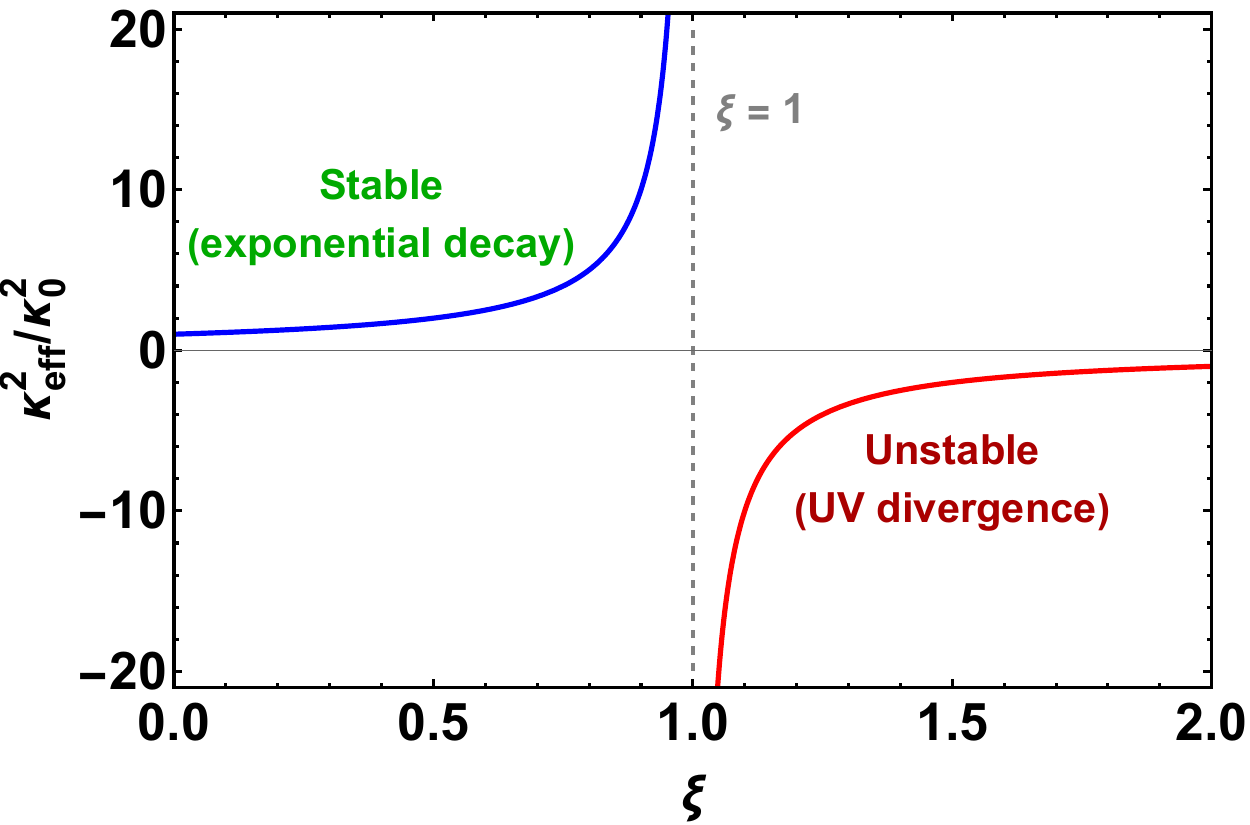}
    \caption{\textbf{Effective screening wave vector squared as a function of the coupling parameter.} 
The normalized squared screening wave vector $\kappa_{\mathrm{eff}}^2/\kappa_0^2 = 1/(1-\xi)$ diverges at the critical coupling $\xi=1$. 
For $\xi<1$ (blue curve), $\kappa_{\mathrm{eff}}^2>0$, corresponding to standard exponential decay of correlations.
For $\xi>1$ (red curve), $\kappa_{\mathrm{eff}}^2<0$, signaling that the effective screening wave vector becomes purely imaginary. Formally, in the minimal continuum model, this signals the dominance of electrostatic attraction over elastic restoring forces; in real colloidal crystals, the discrete lattice cutoff and higher-order elastic terms regulate this ultraviolet divergence.}
    \label{fig:kappa_vs_xi}
\end{figure}

\section{Effective Screening Wave Vector and Structural~Instability}

From the condition $\det\mathbf{M}=0$, we obtain the effective screening wave vector squared
\begin{equation}\label{eq:kappa_final}
\kappa_{\text{eff}}^2 = \frac{\kappa_0^2}{1 - \xi}.
\end{equation}
This result matches the mean-field linearized theory~\cite{Wu2025,Wu2024EPL}. The~behavior of $\kappa_{\text{eff}}^2$ is summarized in Figure~\ref{fig:kappa_vs_xi}:
\begin{itemize}
    \item For $\xi < 1$, $\kappa_{\text{eff}}^2 > 0$, and~the correlation function decays exponentially as $e^{-\kappa_{\text{eff}} r}$.
    \item At $\xi = 1$, $\kappa_{\text{eff}}^2$ diverges, indicating a breakdown of the homogeneous screening picture.
    \item For $\xi > 1$, $\kappa_{\text{eff}}^2 < 0$, so $\kappa_{\text{eff}}$ becomes purely imaginary. Formally, this implies that the poles of the static response function migrate to the real axis, a~signature of a spinodal instability in equilibrium field theory. The~electrostatic attraction overwhelms the elastic restoring forces, driving the uniform phase toward structural collapse at short~wavelengths.
\end{itemize}

\section{Stability Diagram}

\begin{figure}[tbp]
    \centering
    \includegraphics[width=0.48\textwidth]{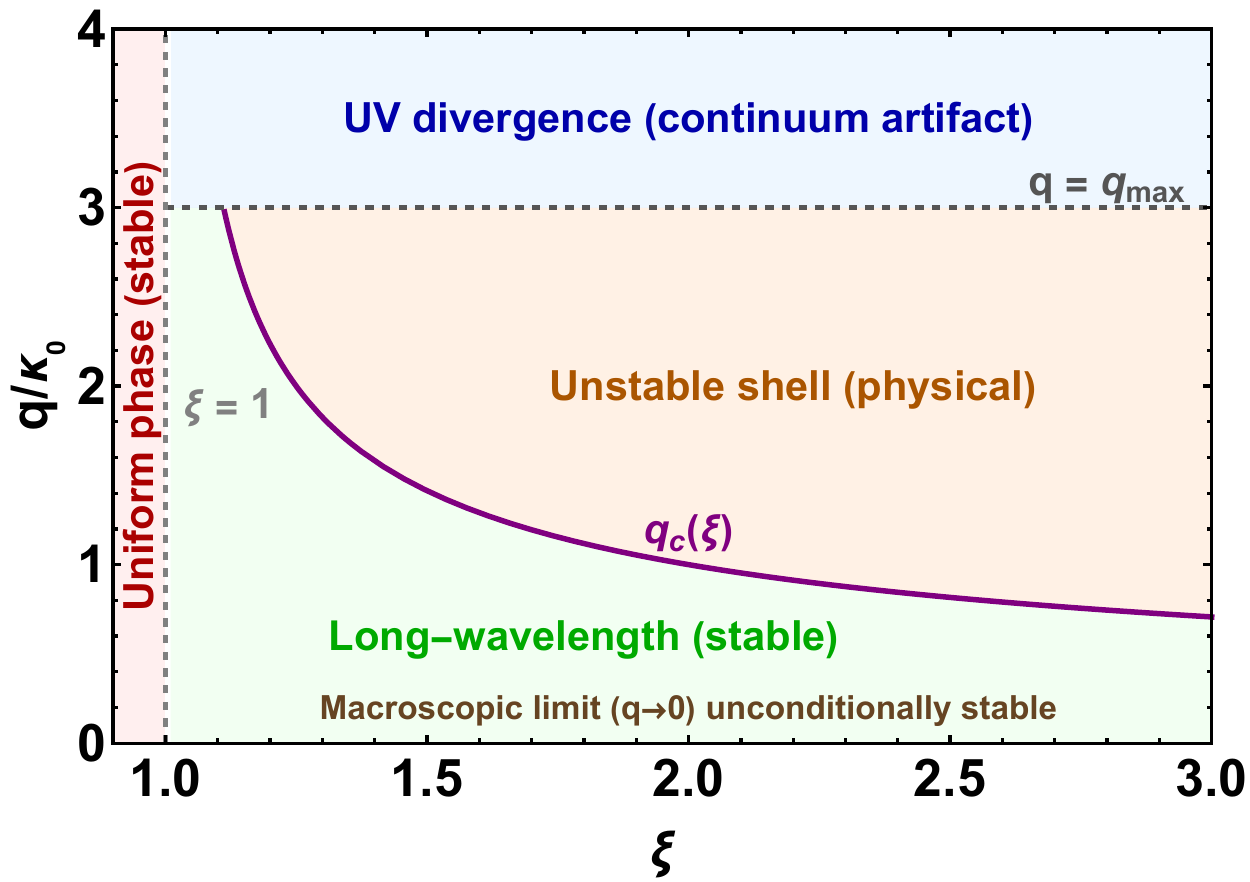}
    \caption{\textbf{Stability diagram of the uniform phase in the minimal electrostatic-elastic coupling model.}
The light red region ($\xi<1$) corresponds to the stable uniform phase. The vertical gray dashed line marks the critical coupling $\xi=1$, where the short-wavelength elastic modulus vanishes. For $\xi>1$, the purple solid curve shows $q_c = \kappa_0/\sqrt{\xi-1}$, above which the uniform phase becomes unstable. The light blue region ($q>q_c$) indicates the continuum prediction of instability extending to arbitrarily short wavelengths. In a real colloidal crystal, the discrete lattice provides an ultraviolet cutoff $q_{\max}\sim\pi/a$; the physically relevant unstable band is the light orange shaded region $q_c < q < q_{\max}$. The horizontal dashed line $q_{\max}$ represents the physical Brillouin zone cutoff, beyond which the continuum model is regularized by the lattice constant. The long-wavelength modes $q<q_c$ (light green region) remain unconditionally stable for all $\xi$, a direct consequence of the perfect ionic screening that protects the macroscopic bulk modulus $\Gamma(0)=\beta K$. Note that the instability only becomes physically manifest when $\xi$ is large enough such that $q_c < q_{\max}$.}
    \label{fig:phase_diagram}
\end{figure}

Figure~\ref{fig:phase_diagram} summarizes the stability regions in the $(\xi,q)$ plane. For~$\xi<1$, the~uniform phase is stable (light red region). At~$\xi=1$, the~short-wavelength modulus $b$ vanishes, and~the system becomes marginally stable at $q\to\infty$. For~$\xi>1$, the~uniform phase is unstable for all wave vectors $q>q_c$ (light blue region). In~the minimal continuum model, this unstable region extends to $q\to\infty$; however, in~a real crystal the discrete lattice imposes a maximum wave vector $q_{\max}$, confining the physical instability to a finite shell $q_c < q < q_{\max}$ (light orange shading). The~long-wavelength modes $q<q_c$ (light green region) remain linearly stable even for $\xi>1$, reflecting the perfect ionic screening that protects the macroscopic bulk modulus $\Gamma(0)=\beta K$. The~light orange region thus represents the physically relevant band of unstable modes, while the unphysical ultraviolet tail is an artifact of the continuum approximation. Note that the instability only becomes physically manifest when $\xi$ is large enough such that $q_c < q_{\max}$.

\unskip

\section{Discussion and~Outlook}

We have shown that the minimal electrostatic--elastic coupling model induces a wave-vector-dependent softening of the effective elastic modulus. The~true long-wavelength modulus $\Gamma(0)$ remains identically equal to the bare modulus $\beta K$, protected by perfect ionic screening at macroscopic scales. In~contrast, the~short-wavelength modulus \mbox{$b = \beta K(1-\xi)$} softens as the coupling $\xi$ increases and vanishes at $\xi=1$, triggering an ultraviolet instability for $q>q_c$. The~absence of higher-order gradient terms in the elastic energy leads to an ultraviolet divergence of the instability, which in real systems is regulated by the discrete lattice cutoff. This indicates that the minimal model correctly captures the onset of local mechanical failure but cannot describe the detailed structure of the resulting phase—whether it is a collapsed state, a~microphase-separated structure, or~a periodic~modulation.

Future work should incorporate strain-gradient terms $\frac{\kappa}{2}(\nabla\theta)^2$ to penalize short-wavelength deformations and to allow for a true Brazovskii-type modulated phase~\cite{Brazovskii1975,Swift1977,Hohenberg1995}. Renormalization-group calculations and numerical simulations will be essential to determine the ultimate fate of the system beyond the instability threshold. The~connection to critical Casimir forces~\cite{Maciolek2018,Gambassi2024} also suggests that near-critical colloidal crystals may exhibit long-range fluctuation-induced interactions, an~intriguing prospect for future experimental and theoretical~studies.

\begin{acknowledgments}
H.W. gratefully acknowledges inspiring discussions with Rudolf Podgornik during the early stages of this work, which profoundly shaped the theoretical framework presented here. This article is dedicated to his memory, in~recognition of his pioneering contributions to the statistical physics of Coulomb systems and his generous mentorship to the soft matter~community. H.W. is supported by the General Program of National Natural Science Foundation of China (NSFC) under Grant No. 12374210, the~open research fund of Songshan Lake Materials Laboratory No. 2023SLABFN20 and the startup fund No. WIUCASQD2022005 from the Wenzhou Institute University of Chinese Academy of Sciences. Z.C. O.Y. is supported by the Major Program of NSFC under Grant No.~22193032.
\end{acknowledgments}

\section*{Data Availability Statement}
No data are available for this theoretical research.

\section*{Conflicts of Interest}
The authors declare no conflicts of interest.

\end{document}